\documentclass[a4paper,dvipdfmx,12pt]{article}
\usepackage{epsf}
\usepackage{color}
\usepackage{cite}
\usepackage{amsmath,amssymb}
\usepackage[dvips]{graphicx}
\usepackage{comment}
\begin{document}

\newcommand{\lsim}{\stackrel{<}{_\sim}}
\newcommand{\gsim}{\stackrel{>}{_\sim}}

\newcommand{\rem}[1]{{$\spadesuit$\bf #1$\spadesuit$}}
\newcommand{\TT}[1]{\textcolor{red}{#1}}

\renewcommand{\theequation}{\thesection.\arabic{equation}}
\renewcommand{\thefootnote}{\fnsymbol{footnote}}
\setcounter{footnote}{0}
\def\thefootnote{\fnsymbol{footnote}}

\begin{titlepage}

\begin{center}

\vskip .75in

{\Large \bf  Impact of EDGES 21cm Global Signal  \vspace{2mm} \\ on Primordial Power Spectrum}
\vskip .75in

{\large
Shintaro Yoshiura$\,^{a}$, Keitaro Takahashi$\,^{a}$, and Tomo~Takahashi$\,^{b}$
}

\vskip 0.25in

{\em
$^{a}$Faculty of Advanced Science and Technology, Kumamoto University, Kumamoto, Japan
\\
$^{b}$Department of Physics, Saga University, Saga 840-8502, Japan
}

\end{center}
\vskip .5in

\begin{abstract}

We investigate the impact of the recent observation of the 21cm global signal by EDGES on primordial power spectrum,
particularly focusing on the running parameters $\alpha_s$ and $ \beta_s$ which characterize the detailed scale dependence of the primordial spectrum.
When primordial power spectrum is enhanced/suppressed on small scales, the structure formation proceeds faster/slower and changes the abundance of 
small size halos, which affects the sources of Lyman $\alpha$ radiation at high redshifts  to alter the position of the absorption line.
Recent observation of EDGES detected the 21cm absorption line at $z=17.2$ and this result also indicates that 
the brightness temperature  is consistent with zero for $z \lesssim 14$ and  $z \gtrsim 22$, 
which can exclude a scenario giving the absorption line at such  redshifts. 
We argue that the bound on the running parameters can be obtained by requiring that the absorption line 
should not exceed observational bounds at such  redshift ranges and found that  some parameter space of $\alpha_s$ and $\beta_s$ allowed by Planck may be disfavored
for some values of astrophysical parameters.

\end{abstract}

\end{titlepage}

\renewcommand{\thepage}{\arabic{page}}
\setcounter{page}{1}
\renewcommand{\thefootnote}{\#\arabic{footnote}}
\setcounter{footnote}{0}

%%%%%%%%%%%%%%%%%%%%%%%%%%%%%%%%%%%%%%%%%%%%%%%%%%
\section{Introduction} \label{sec:intro}
%%%%%%%%%%%%%%%%%%%%%%%%%%%%%%%%%%%%%%%%%%%%%%%%%%

The nature of primordial fluctuations reflects the physical mechanism behind the inflationary Universe and 
hence a great deal of effort has been made to understand it both from observational and theoretical viewpoints.
Current cosmological observations such as  cosmic microwave background (CMB) from Planck satellite  have measured 
primordial power spectrum, in particular, its amplitude and the spectral index $n_s$ very accurately \cite{Ade:2015xua,Ade:2015lrj}, 
and  together with the bound on the amplitude of primordial gravitational waves such as from CMB B-mode observations like BICEP2/KECK \cite{Array:2015xqh}, 
the inflationary models are now severely tested. Nevertheless there remains  a large variety of inflationary  models consistent with those observations 
and we are still far from  a thorough understanding of the inflationary Universe. 
Therefore it is imperative to go further to probe primordial fluctuations. 

One direction would be to measure the primordial power spectrum more precisely. 
For this purpose, it is common to adopt the following parametrization for the primordial spectrum:
\begin{equation}
%\label{ }
P_\zeta (k) = A_s (k_{\rm ref}) \left( \frac{k}{k_{\rm ref}} \right)^{n_s- 1 + \frac12 \alpha_s  \ln (k/k_{\rm ref}) + \frac{1}{3!} \beta_s   \ln^2 (k/k_{\rm ref})  },
\end{equation}
where $A_s (k_{\rm ref}) $ is the amplitude at the reference scale $k_{\rm ref}$, $n_s$ is the spectral index.  $\alpha_s$ and $\beta_s$ are 
the so-called running parameters and represent the detailed scale dependence of $P_\zeta (k)$ as
\begin{equation}
%\label{ }
\alpha_s = \left. \frac{d^2  \ln P_\zeta (k)}{d \ln k^2}\right|_{k=k_{\rm ref}} \, ,
\qquad
\beta_s = \left. \frac{d^3  \ln P_\zeta (k)}{d \ln k^3}\right|_{k=k_{\rm ref}} \, .
\end{equation}
Although one can obtain the bounds on the running parameters from Planck data, they are not so severe  and 
thus it is worth exploring a possibility of probing the runnings by using yet another observation to measure them more accurately. 

When one wishes to determine the runnings precisely,  probing the power spectrum  on a wide range of scales would be helpful.
Since CMB measures cosmic fluctuations on large scales, one needs observations on small scales.
As such,  expected constraints from future observations of 21cm fluctuations of neutral hydrogen from intergalactic 
medium \cite{Kohri:2013mxa,Munoz:2016owz} and minihalos \cite{Sekiguchi:2017cdy}, 
and CMB spectral $\mu$ distortions\footnote{
In addition to the $\mu$ distortion \cite{Sunyaev:1970er,Hu:1992dc,Chluba:2011hw,Chluba:2012gq,Khatri:2012rt}, the $i$-distortion \cite{Khatri:2012tw} can also give a useful information on primordial power spectrum.}
\cite{Dent:2012ne,Khatri:2013dha,Cabass:2016giw,Kainulainen:2017gqq} have been studied. 

We in this paper investigate how the nature of primordial power spectrum affects the 21cm global signal
and the impact of the recent EDGES result \cite{Bowman:2018yin} on the running parameters. 
EDGES has reported that the absorption peak is observed at the frequency corresponding to $z=17.2$ and its brightness temperature relative to CMB 
is $ T_b = -500^{+200}_{-500}~{\rm mK}~{\rm (99\%~C.L.)}$, whose value is too low to be explained in the standard cosmological and astrophysical 
scenarios\footnote{
See, however,  Ref.~\cite{Hills:2018vyr} for the discussion on the analysis   and its interpretation.
}. This result has stimulated many works which aim to explain its non-standard value by resorting to dark matter (DM) interactions to cool the gas (baryon) \cite{Barkana:2018lgd,Munoz:2018pzp,Fialkov:2018xre,Kang:2018qhi,Falkowski:2018qdj,Lambiase:2018lhs,Jia:2018csj,Sikivie:2018tml}, 
which could be confronted with other cosmological/astrophysical observations \cite{Berlin:2018sjs,Barkana:2018qrx,Slatyer:2018aqg}. 
Other possibilities to explain the signal have been discussed in models with producing photons at 
radio wavelength \cite{Ewall-Wice:2018bzf,Fraser:2018acy,Yang:2018gjd,Pospelov:2018kdh,Lawson:2018qkc,Moroi:2018vci,Aristizabal:2018},  dark sector properties  \cite{Costa:2018aoy,Hill:2018lfx,Li:2018kzs} and so on. On the other hand, there have been  several works to derive constraints on DM and 
primordial black holes by using the EDGES result \cite{DAmico:2018sxd,Safarzadeh:2018hhg,Clark:2018ghm,Cheung:2018vww,Hektor:2018qqw,Liu:2018uzy,Mitridate:2018iag,Schneider:2018xba,Lidz:2018fqo,Hektor:2018}. 
Implications of the EDGES result for 21cm power spectrum have also been investigated \cite{Munoz:2018jwq,Kaurov:2018kez}.

One should  notice that the EDGES result also indicates that there is no absorption signal at the redshift regions $z  \lesssim  14$ and $z \gtrsim 22$,
which can  constrain the running parameters of primordial power spectrum.
Since the runnings $\alpha_s$ and $\beta_s$ directly affect the matter power on small scales, the structure formation should change depending on these parameters. 
For too positively (negatively) large values of  $\alpha_s$ and $\beta_s$, the matter power on small scales are enhanced (suppressed) and the structure formation  is  affected. 
When the structure formation proceeds faster (slower) due to the enhanced (suppressed) matter power spectrum on small scales, 
which switches on the source of Lyman $\alpha$ radiations earlier (later), then the absorption line is shifted to a higher (lower) redshift\footnote{
When the running parameters are too negative where  the matter power on small scales is suppressed, its effects are similar to the case with 
warm dark matter \cite{Sitwell:2013fpa,Schneider:2018xba}.
}. 
If the absorption line appears at the redshift ranges $z \lesssim 14$ or $z \gtrsim 22$, it is inconsistent with the EDGES result and such parameter values are disfavored. By adopting this argument, 
we can obtain  the bound on the running parameters  $\alpha_s$ and $\beta_s$ from EDGES, which is the main purpose of this paper. 

The organization of this paper is the following. In the next section, we briefly describe the procedure to compute the 21cm brightness temperature and
discuss the effects of the runnings on the global signal of 21cm line.
Then in Section~\ref{sec:bound}, we investigate the bound on the running parameters from EDGES by requiring that 
the absorption line should not appear at the redshift ranges $z \lesssim 14$ and $z \gtrsim 22$.
The effects of astrophysical parameters on the 21cm global signal and constraints on the running parameters are discussed in Section~\ref{sec:astro_param}.
The final section is devoted to conclusion of this paper.

%%%%%%%%%%%%%%%%%%%%%%%%%%%%%%%%%%%%%%%%%%%%%%%%%%
\section{21cm global signal and primordial power spectrum} \label{sec:global_signal_primordial}
%%%%%%%%%%%%%%%%%%%%%%%%%%%%%%%%%%%%%%%%%%%%%%%%%%

Here we  briefly discuss  how the nature of primordial power spectrum  affects the global signal of 
21cm line of neutral hydrogen.  

The 21cm global signal,  averaged differential brightness temperature relative to CMB radiation,  is given by \cite{Wouthuysen:1952,Field:1958,Field:1959} (see, e.g., the reviews \cite{Furlanetto:2006jb,Pritchard:2011xb})
\begin{equation}
\label{eq:T_b}
T_b \simeq 27\, x_{\rm HI} \left( \frac{\Omega_bh^2}{0.023} \right) \left( \frac{0.15}{\Omega_m h^2} \right)^{1/2}  \left( \frac{1+z}{10} \right)^{1/2} 
\left( \frac{T_s - T_{\rm CMB}}{T_s} \right)\, {\rm mK}\,  , 
\end{equation}
where $\Omega_b, \Omega_m$ are density parameters for baryon and (total) matter, $h$ is the  Hubble parameter in units of $100\, {\rm km}\, {\rm s}^{-1}\, {\rm Mpc}^{-1}$,
$T_s$ is the spin temperature of neutral hydrogen, $T_{\rm CMB} = 2.725 /(1+z)$ is the  temperature of CMB photon.
The spin temperature $T_s$ can be given by 
\begin{equation}
%\label{ }
T_s^{-1} = \frac{ T_{\rm CMB}^{-1} + x_\alpha T_\alpha^{-1} + x_c T_K^{-1}}{1 + x_\alpha + x_c}\, ,
\end{equation}
where $T_\alpha$ and $T_K$ respectively are  the color temperature and gas temperature, respectively. 
$x_\alpha$ and $ x_c$ are the coupling coefficients characterizing  the scattering of Lyman $\alpha$ photons and the atomic collisions.
For the detailed discussion of the evolution of the spin temperature, see e.g.,  Refs.~\cite{Furlanetto:2006jb,Pritchard:2011xb}.
 
To investigate how the running parameters $\alpha_s$ and $\beta_s$ affect  the brightness temperature, 
we have computed the evolution of the spin temperature  by using 21cmFAST \cite{Mesinger:2007pd,Mesinger:2010ne}.
To calculate its evolution, we also need to specify astrophysical parameters such as the minimum virial temperature of star forming sources $T_{\rm vir}$, the fraction of baryons converted to stars $f_\ast$ 
and the number of X-ray photons emitted from stars $\zeta_{\rm X}$.
The evolution of the spin temperature is also sensitive to these astrophysical parameters\footnote{
{In 21cmFAST, the evolution of the spin temperature is solved with semi-analytic models and approximations such as the one regarding the characterization of X-ray sources and a step function for the optical depth. The accuracy of ionization model has been investigated in \cite{Mesinger:2010ne,Zahn2011}, and the power spectrum matches with the numerical simulation within 10\% of error. However, the accuracy of 21cmFAST for the global signal has not been discussed, and one has to perform a massive radiative transfer simulation and 21cmFAST with the same initial condition to validate its accuracy.  This is outside the scope of this paper. However, 
it should be mentioned that the entire evolution of the global signal is fairly consistent with the results obtained by using independent semi-numerical simulations (e.g. \cite{Santos2010MNRAS,Fialkov2014MNRAS}) and radiative transfer simulations (e.g. \cite{Ross2017MNRAS,Semelin2017MNRAS}).}
}. We discuss how these parameters affects the absorption line in Section~\ref{sec:astro_param}.

\begin{figure}[htbp]
\begin{center}

%\resizebox{130mm}{!}{
\hspace{-30mm}
     \includegraphics[width=7cm,bb=50 0 330 272]{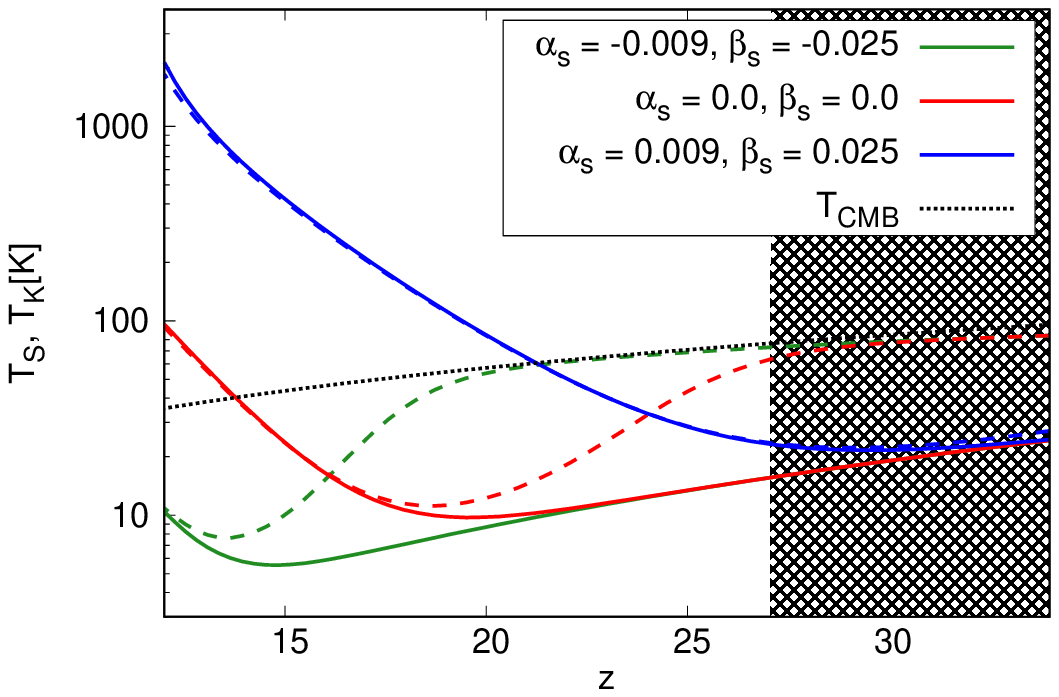}
     \vspace{-10mm}
\hspace{10mm}
    \includegraphics[width=7cm,bb=50 0 330 272]{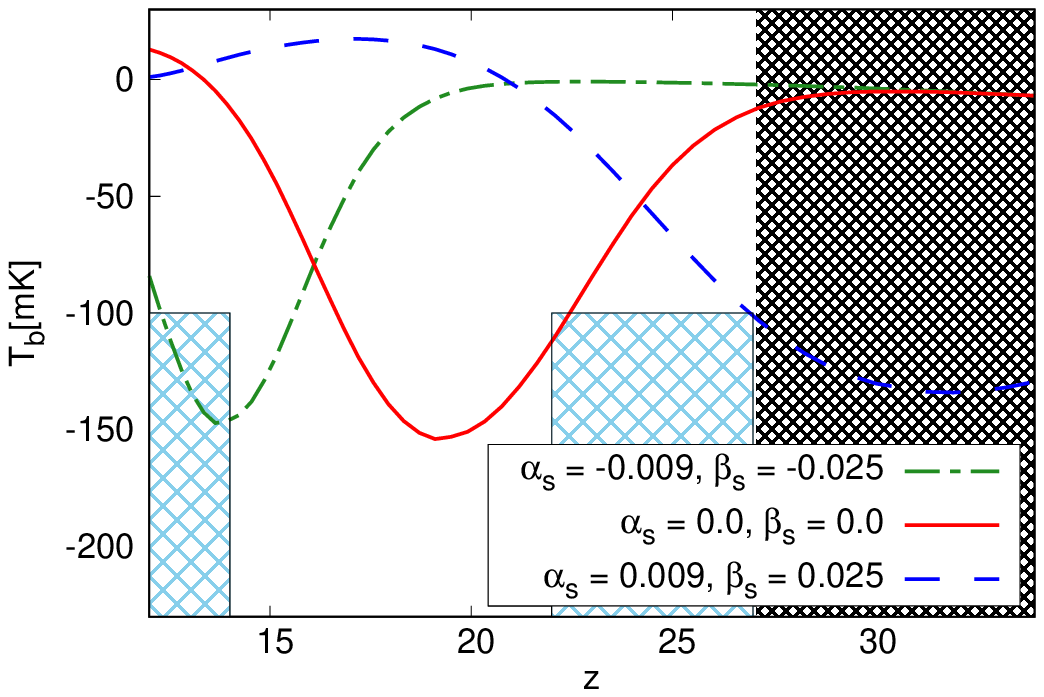}
    \vspace{-10mm}
%}
\end{center}
\caption{[Left panel]:  Evolution of $T_k$ (solid), $T_s$ (dashed) and $T_{\rm CMB} $ (dotted). 
  [Right panel]: Evolution of the brightness temperature $T_b$. 
 Cases with $(n_s, \alpha_s, \beta_s) = (0.9586, 0, 0)$ (red), $(0.9586, 0.009, 0.025)$ (blue) and 
  $(0.9586, -0.009, -0.025)$ (green) are shown. Light blue hatched regions correspond to the ones inconsistent with the EDGES result.
  Black hatched one indicates that there is no data in the redshift range.}
\label{fig:Ts_Tb}
\end{figure}

In Fig.~\ref{fig:Ts_Tb}, we show the evolutions of $T_s, T_K$  and $T_{\rm CMB}$ (left panel) and the differential brightness temperature $T_b$ (right panel)
for several parameter sets of $(n_s, \alpha_s, \beta_s)$, whose values are indicated in the figure.  The astrophysical parameters 
mentioned above are assumed as $ T_{\rm vir} =10^4~{\rm K}, f_\ast = 0.05$ and $\zeta_X =2\times 10^{56} /M_\odot$.
The cosmological parameters are taken as follows: $\sigma_8 = 0.831, \Omega_b h^2 = 0.02225, \Omega_m = 0.3156, h = 0.6727$
where $\sigma_8$ is the amplitude of matter fluctuations at $8 h^{-1}$~Mpc.  As mentioned in the introduction, when $\alpha_s$ and $\beta_s$ are positively (negatively) large, 
the matter power on small scales are enhanced (suppressed), and thus the structure formation proceeds faster (slower) and the sources of Lyman $\alpha$ background are switched on 
earlier (later). Therefore positively (negatively) larger values of  $\alpha_s$ and $\beta_s$  drive the spin temperature to approach the gas temperature earlier (later), which shifts the absorption line to a higher (lower) redshift. 
On the other hand, the recent EDGES result indicates that the absorption line appears at around $z = 17.2$ and the brightness 
temperature goes to zero for  $z  \lesssim 14$ and $z \gtrsim 22$, therefore too large positive or negative values of the runnings are expected to be disfavored in the light of recent 
EDGES result.  In the next section, we investigate the bound on the runnings $\alpha_s$ and $\beta_s$ from the requirement  that 
the redshift (frequency) of the absorption line should be consistent with the EDGES result, i.e., it should not appear at  $z \lesssim 14$  and $z \gtrsim 22$.

%%%%%%%FIGURE%%%%%%%%%%%%%%%%%%%%%%
%%%%%%%%%%%%%%%%%%%%%%%%%%%%%%%%%

%%%%%%%%%%%%%%%%%%%%%%%%%%%%%%%%%%%%%%%%%%%%%%%%%%
\section{Bounds on primordial power spectrum from EDGES 21cm global signal} \label{sec:bound}
%%%%%%%%%%%%%%%%%%%%%%%%%%%%%%%%%%%%%%%%%%%%%%%%%%

Now in this section, we study the bound on the running parameters $\alpha_s$ and $\beta_s$ of primordial 
power spectrum by demanding that the absorption line should appear in the redshift range of $14< z < 22$ indicated by the EDGES result. 
In practice, we conservatively require that the brightness temperature should be  $T_b > -100~{\rm mK}$ for the redshift range 
 $z < 14$ and $z > 22$. We note that, since EDGES does not give data for $z \gtrsim 27$, we do not constrain the case where 
 the absorption trough appears at redshifts $z>27$.

\begin{figure}
\begin{center}
\vspace{10mm}
\resizebox{170mm}{!}{
\hspace{0mm}
     \includegraphics[width=5cm]{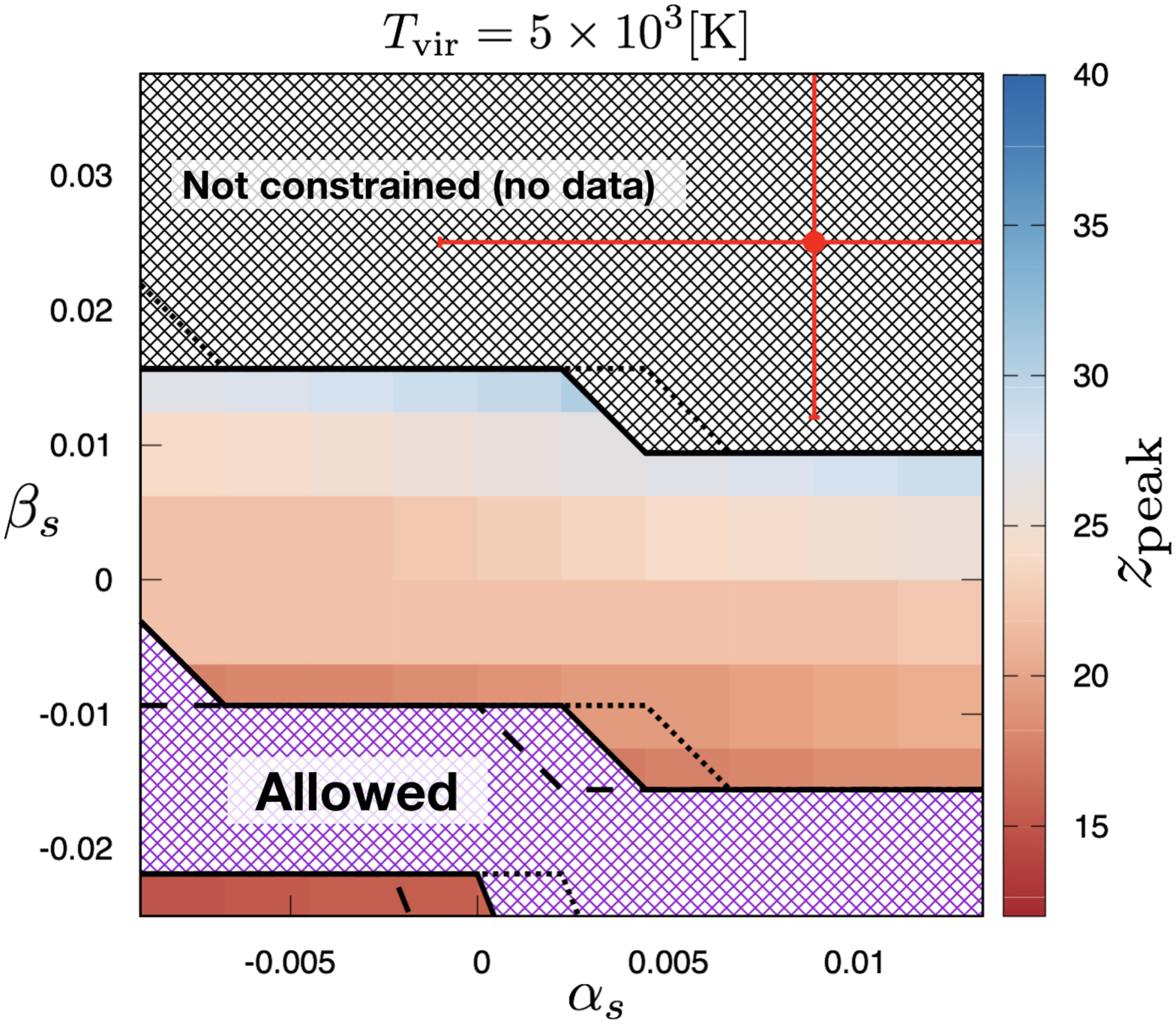}
\hspace{0mm}
     \includegraphics[width=5cm]{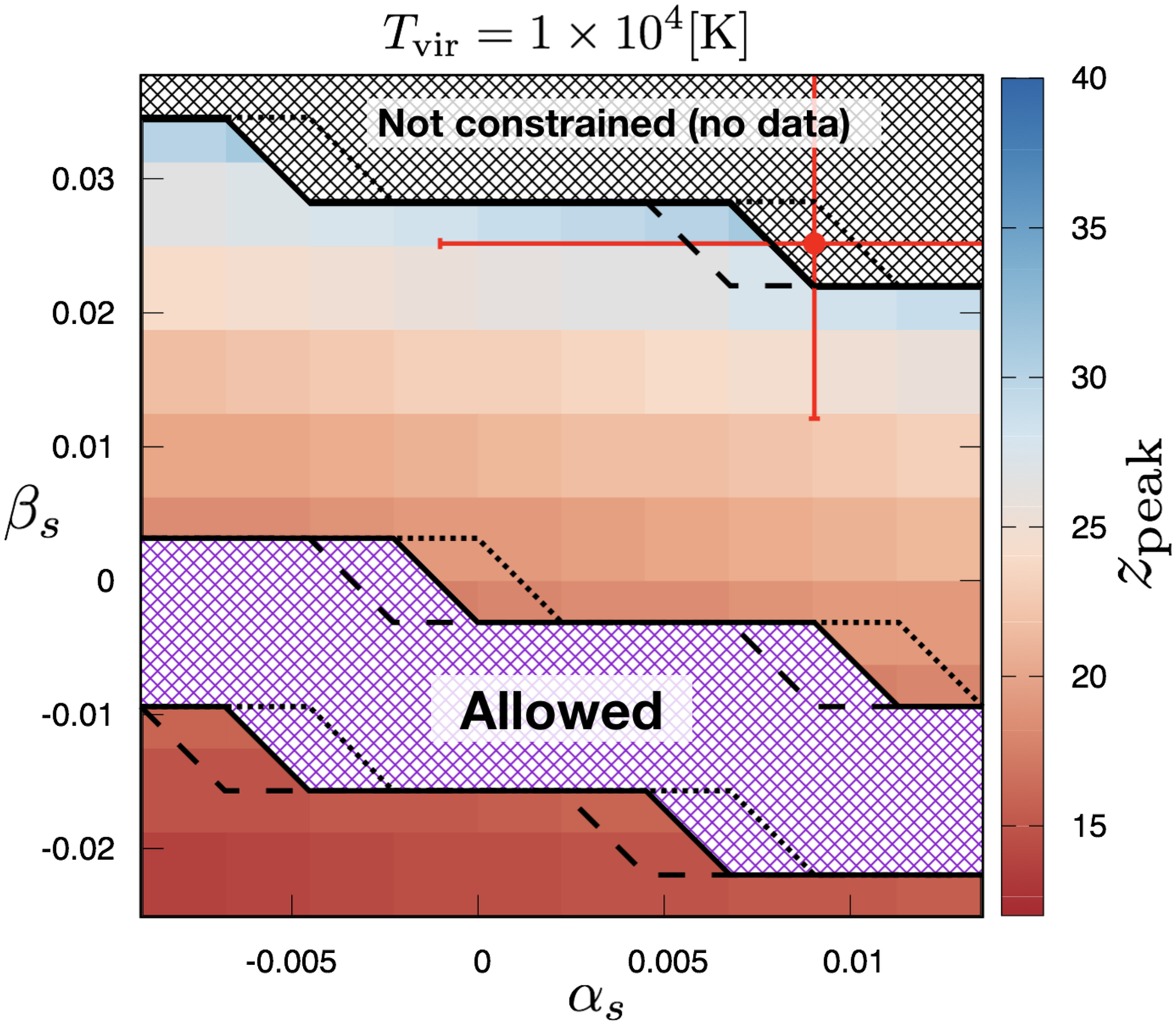}
\hspace{0mm}
     \includegraphics[width=5cm]{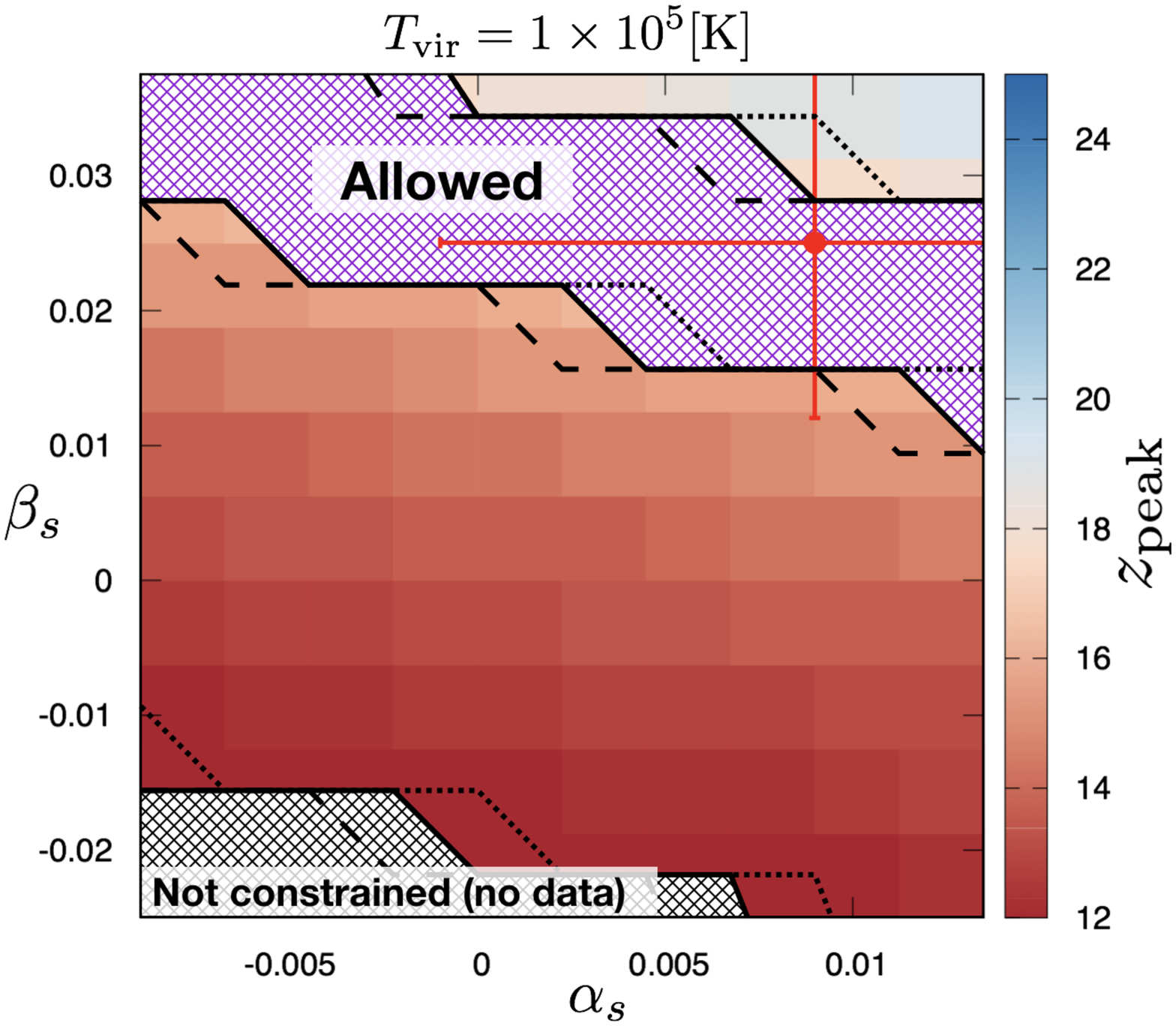}
}
\end{center}
\caption{Allowed region from the EDGES result and the peak redshift of the absorption trough  $z_{\rm peak}$ are shown with a color panel
for the cases with with $T_{\rm vir} = 5\times 10^{3}~{\rm K}$ (left),  $10^4~{\rm K}$ (middle) and $10^5~{\rm K}$ (right). 
Other cosmological and astrophysical parameters are fixed as described in the text.
Here $n_s$ is fixed to be  $n_s = 0.9530, 0.9586$ and $0.9642$ which correspond to 
the values of 1$\sigma$ lower, central and 1$\sigma$ upper bound from Planck and  depicted as dotted, solid and dashed lines, respectively.
Red point shows Planck best fit value with 1 sigma error bars. 
When the peak redshift  of the absorption line is in the range of $z> 27$ or $T_b > -100~{\rm mK}$ at $z=27$, no constraint can be obtained from EDGES, which is indicated by black hatched. 
It should be noted that the constraint depends on the values of astrophysical parameters, which will be  discussed in Section~\ref{sec:astro_param}.
}
\label{fig:bound}
\end{figure}

In Fig.~\ref{fig:bound}, color panels of the redshift of the peak position of the absorption trough  $z_{\rm peak}$ and allowed region  from EDGES are shown
in the $\alpha_s$--$\beta_s$ plane.  To obtain a constraint from EDGES, we basically only use the information of the position of the absorption trough (not of the amplitude).
As  mentioned, the EDGES result indicates that the absorption line should appear in the redshift range of $ 14 < z < 22$.
We define a criterion for the allowed model such that the absorption line fall into this redshift range.
To be precise, for the allowed model, we require that the predicted evolution of $T_b$ does not cross the blue hatched regions in the right panel of Fig.~\ref{fig:Ts_Tb},
where we conservatively demand that $T_b > -100~{\rm mK}$ even for $z < 14$ and $z>22$, taking observational errors into account.
Since there is no data for $z>27$ and $z<13.2$, when the peak of the absorption trough is $z>27$ or $z<13.2$, we cannot obtain any constraint, which is 
shown as ``Not constrained (no data)" in the figure.
We consider a flat $\Lambda$CDM model and fixed other cosmological parameters as $\sigma_8 = 0.831, \Omega_b h^2 = 0.02225, \Omega_m = 0.3156, h = 0.6727$
as  assumed in Fig.~\ref{fig:Ts_Tb} since these parameters are well measured by Planck \cite{Ade:2015xua}. 
By doing so in the analysis, we effectively combine the Planck results with the 21cm global signal by EDGES. 
Regarding the spectral index $n_s$, its effect on the brightness temperature is degenerate with the running parameters $\alpha_s$ and $\beta_s$, 
and hence  we take $n_s = 0.9530, 0.9586$ and $0.9642$ which are respectively the values of 1$\sigma$ lower, central and 1$\sigma$ upper bound from Planck  \cite{Ade:2015lrj},
and respectively depicted with dotted, solid and dashed lines in Fig.~\ref{fig:bound}. As seen from the figure, the change of $n_s$ within the 1$\sigma$ range allowed by Planck 
does not affect the constraint on $\alpha_s$ and $\beta_s$ much.
For astrophysical parameters, we take $T_{\rm vir} = 5\times 10^{3}~{\rm K}$ (left),  $10^4~{\rm K}$ (middle) and $10^5~{\rm K}$ (right).
Other astrophysical parameters are assumed as $ f_\ast = 0.05$ and $\zeta_X =2\times 10^{56} /M_\odot$. With this value of $\zeta_X$, 0.3 X-ray photons are emitted per stellar baryon.

%%%%%%%%%%%%%%%%%%%%%%%%%%%%%%%%%%
%%%%%%%%%%%%%%%%%%%%%%%%%%%%%%%%%%

Since large positive and negative values of $\alpha_s$ and $\beta_s$ give too high or too low redshift of the absorption trough as discussed in the previous section, 
such regions can be considered to be excluded by the EDGES result. 
We note that the constraints on $\alpha_s$ and $\beta_s$ derived from Planck 2015 TT, TE, EE+lowP analysis are \cite{Ade:2015lrj}
\begin{equation}
%\label{ }
\alpha_s =  0.009 \pm 0.010\, , 
\qquad
\beta_s = 0.025 \pm 0.013\, ,
\end{equation}
which is also indicated in Fig~\ref{fig:bound}.

Although constraints on $\alpha_s$ and $\beta_s$ depend on the values of astrophysical parameters such as $T_{\rm vir}$ as shown in Fig.~\ref{fig:bound}, 
it is interesting to see that  some parameter region for $\alpha_s$ and $\beta_s$ allowed by Planck at 1$\sigma$ can be  
ruled out by the  EDGES for some values of $T_{\rm vir}$. This shows that the 21cm global signal is powerful  in constraining the runnings of primordial power spectrum
although some of the parameter space cannot be constrained since the EDGES measures the redshift range of $14 \lesssim z \lesssim 27$.
We should also note that, although the recent result from EDGES may need to be confirmed by other observations of the 21cm global signal, 
as far as the frequency range of the absorption trough persists (even if the size of the absorption signal is weakened), the constraint obtained 
in this paper will still be valid. 

As mentioned above,  the constraint on the running parameters depend on the astrophysical parameters such as $T_{\rm vir}$. 
In the next section, we discuss effects of the astrophysical 
parameters on the evolution of the brightness temperature and how the constraint is affected in some detail.

%%%%%%%%%%%%%%%%%%%%%%%%%%%%%%%%%%%%%%%%%%%%%%%%%%
\section{Effects of Astrophysical Parameters} \label{sec:astro_param}
%%%%%%%%%%%%%%%%%%%%%%%%%%%%%%%%%%%%%%%%%%%%%%%%%%

As discussed in the previous section,  our analysis indicates that some parameter space of $\alpha_s$ and $\beta_s$ allowed by 
Planck could be disfavored by the EDGES result  for some fixed astrophysical parameters.
However, the 21cm global signal also strongly depends on the astrophysical parameters used in 21cmFAST. 
Here we discuss in some detail how the astrophysical parameters such as $T_{\rm vir}, f_\ast$ and $\zeta_X$ affect the 21cm absorption signal\footnote{
{Another parameter such as the ionizing efficiency $\zeta$ can also affect the 21cm signal. However, we note that
 the ionization starts after X-ray heating with our choice of the parameters such as the ionizing efficiency $\zeta = 20$ and $\zeta_X = 2 \times 10^{56} /M_\odot$. Furthermore, models with the high ionization rate are ruled out by the constraint of Thomson scattering optical depth of the CMB. Thus, these parameters may not change the results. }
}.

%%%%%%%FIGURE%%%%%%%%%%%%%%%%%%%%%%
\begin{figure}[htbp]
\begin{center}

\vspace{-30mm}

\hspace{-10mm}
\resizebox{100mm}{!}{\includegraphics[width=9cm]{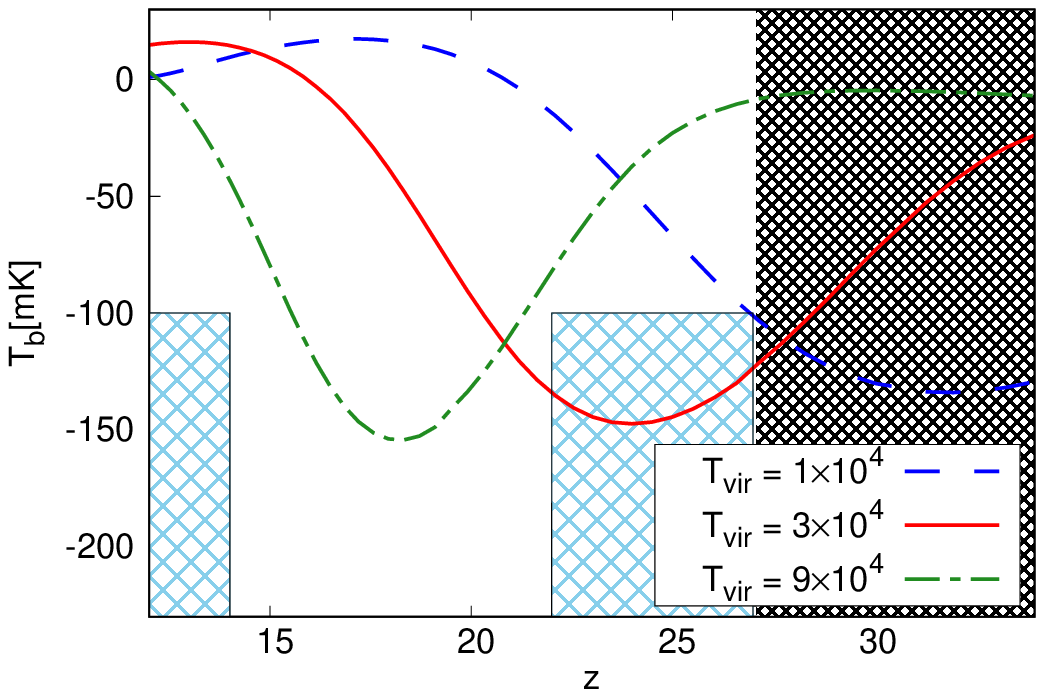}} 

\vspace{-10mm}

\hspace{-10mm}
\resizebox{100mm}{!}{\includegraphics[width=9cm]{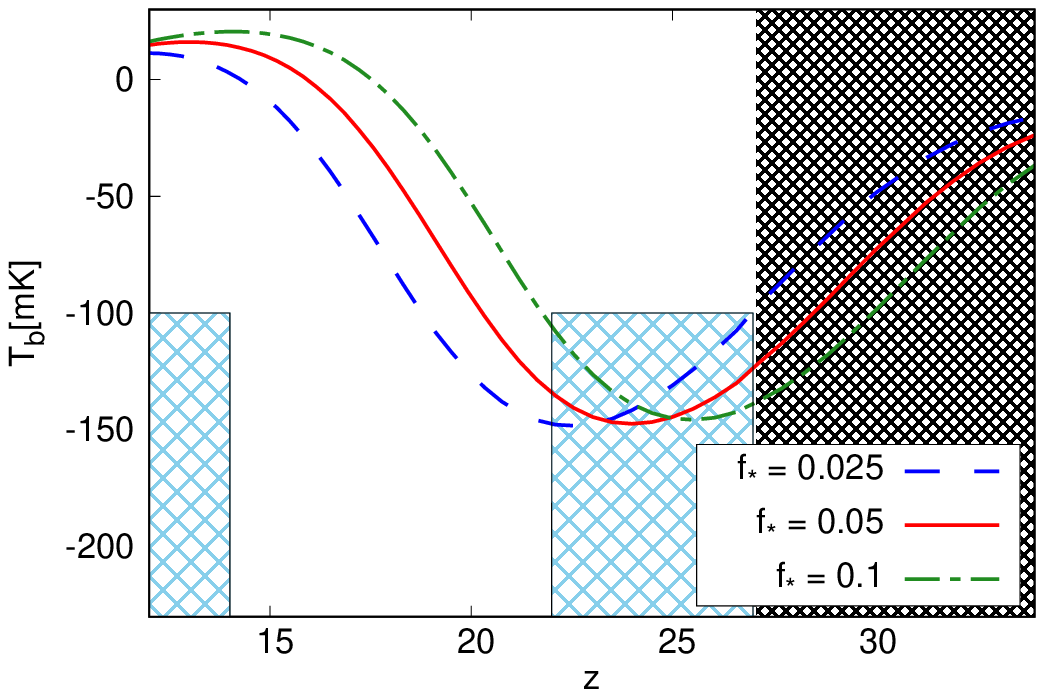}}

\vspace{-10mm}

\hspace{-10mm}
\resizebox{100mm}{!}{\includegraphics[width=9cm]{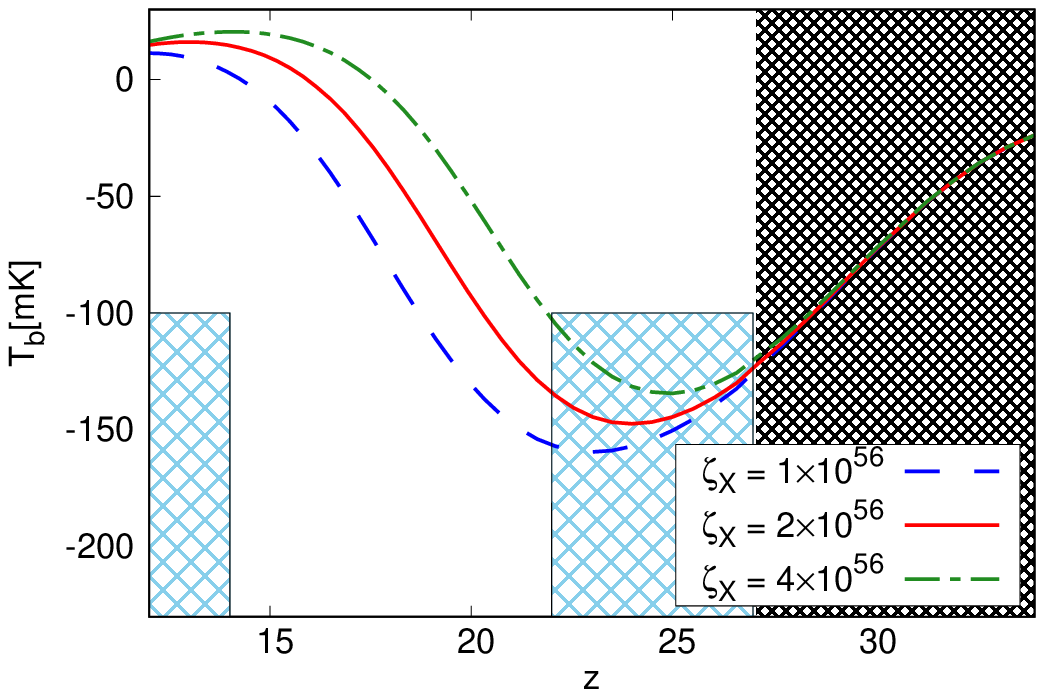}} 
\vspace{-10mm}
\end{center}
\caption{
Evolution of the brightness temperature for various sets of astrophysical parameters. We show the impact of minimum virial temperature, $T_{\rm vir}$ (top), fraction of baryon converted to starts, $f_{\ast}$ (middle) 
and X-ray heating efficiency, $\zeta_{\rm X}$ (bottom). Here we assume $n_s = 0.9586$, $\alpha_s =0.009$, $\beta_s = 0.025$ for the spectral index and its runnings.
Light blue hatched regions correspond to the ones inconsistent with the EDGES result.  Black hatched one indicates that there is no data in the redshift range.
}
\label{fig:GSastro}
\end{figure}
%%%%%%%%%%%%%%%%%%%%%%%%%%%%%%%%%

In Fig.~\ref{fig:GSastro}, we show the evolution of the brightness temperature $T_b$ by varying the values of $T_{\rm vir}$ (top panel), $f_\ast$ (middle panel) and $\zeta_{X}$ (bottom panel). 
Since $T_{\rm vir}$ and $f_\ast$ affect the WF coupling and gas heating in the same way, the effects of these parameters on the 21cm global signal are degenerate.
But on the other hand, $\zeta_X$ only affects gas heating, and hence the variation of $\zeta_X$ is not degenerate with the other parameters.

Concerning $T_{\rm vir}$,  a negative feedback effect such as Lyman-Werner radiation at high redshifts leads to the 
suppression of the star formation in halos \cite{Haiman:1999mn}, which results in a larger value of $T_{\rm vir}$ and the reduction of X-ray and Lyman-$\alpha$ photons from small halos, 
then  the absorption line shifts to a lower redshift.
The fraction of baryon converted to stars $f_\ast$ also alters the number of X-ray and Lyman-$\alpha$ photons. 
Smaller values of $f_\ast$ make the WF coupling less effective at higher redshifts and gas heating occur later, and thus the absorption peak shifts to a lower redshift, 
whose effect is quite similar to the one for $T_{\rm vir}$.  Therefore, as can be seen from Fig.~\ref{fig:GSastro},  the change in $T_{\rm vir}$ and $f_\ast$ give a degenerate effect on the 
evolution of $T_b$.

\begin{figure}%[htbp]
\begin{center}
\vspace{10mm}
\resizebox{160mm}{!}{
\hspace{0mm}
     \includegraphics[width=8cm]{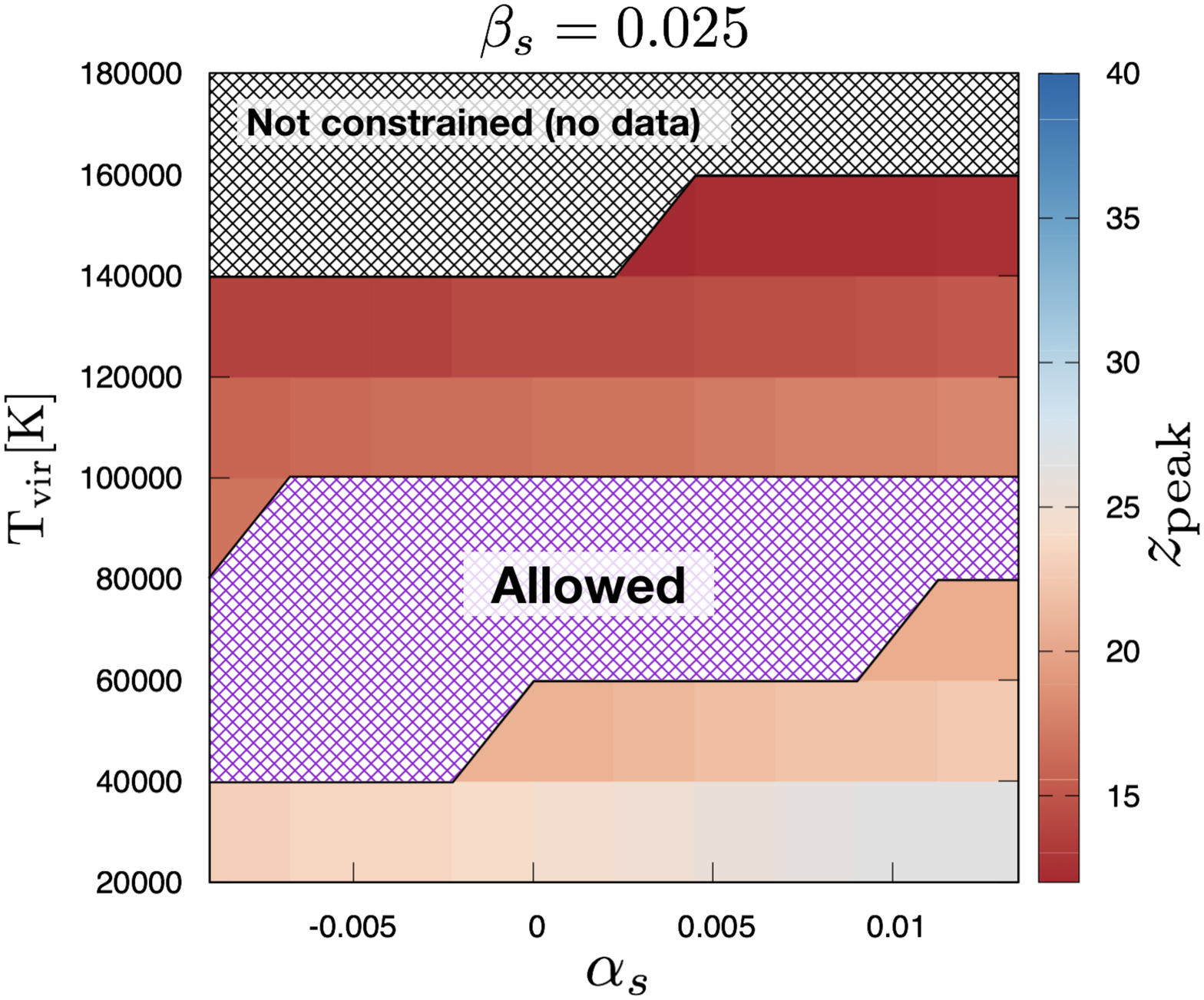}
\hspace{0mm}
     \includegraphics[width=8cm]{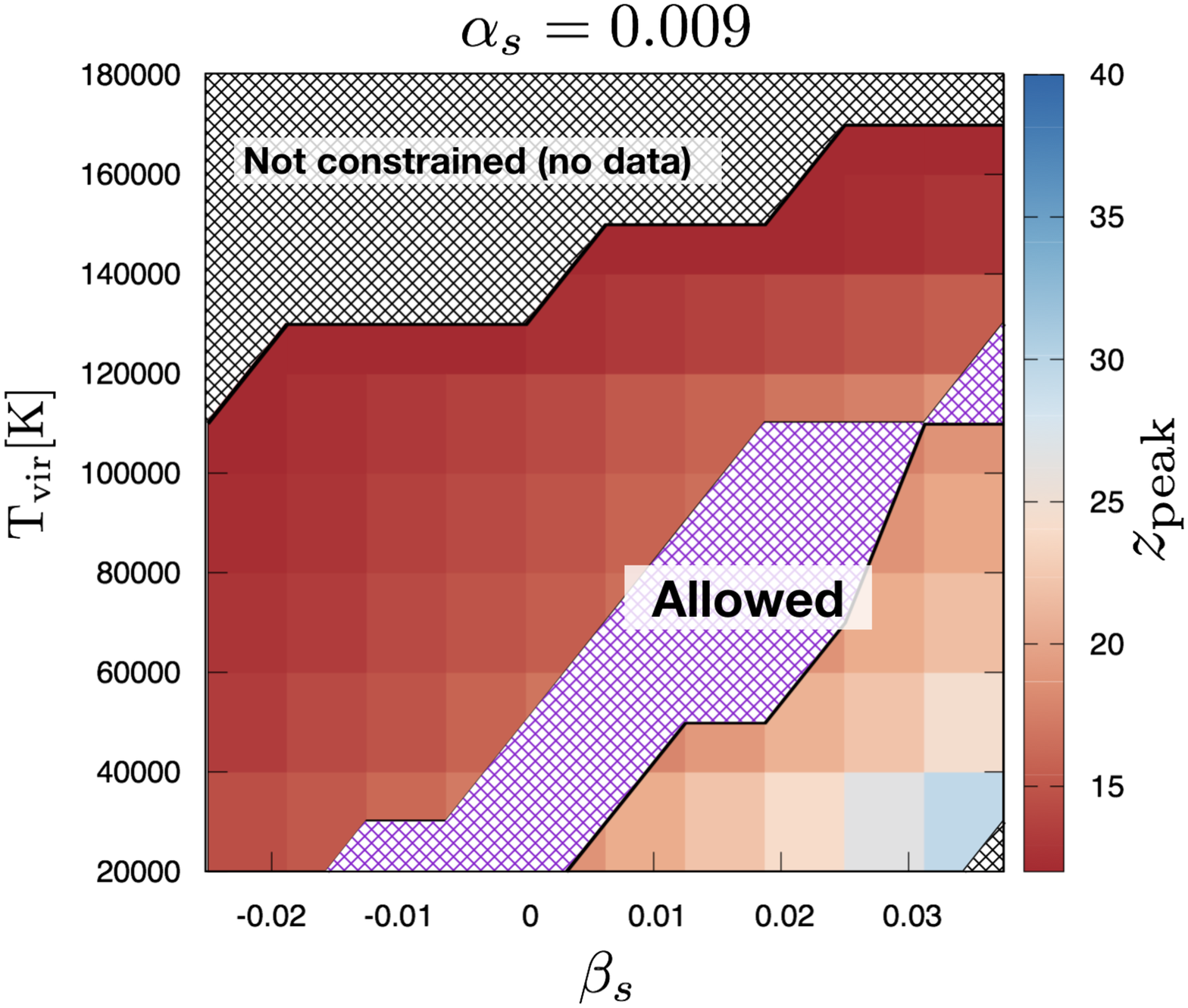}
}
\end{center}
\caption{ Allowed region from the EDGES result and the peak redshift of the absorption trough  $z_{\rm peak}$ are shown with a color panel in the 
$\alpha_s$--$T_{\rm vir}$ (left) and $\beta_s$--$T_{\rm vir}$ (right) planes. For the left and right panels, we fix the other running parameter as $\beta_s = 0.025$ (left) and $\alpha_s = 0.009$ (right).
Other cosmological and astrophysical parameters are fixed as in Fig.~\ref{fig:bound}.
{In the right panel, although the allowed region is disconnected at around $(\beta_s, T_{\rm vir})  \sim (0.03, 1.1 \times 10^5~{\rm [K]})$, this is due to sparse grid sampling in the analysis. 
Since $\beta_s$ and $T_{\rm vir}$ are degenerate, the allowed regions should be connected if we adopt denser sampling.}
}
\label{fig:bound2}
\end{figure}

In Fig.~\ref{fig:bound2}, to see how the values of $T_{\rm vir}$ affects the constraints on the running parameters, 
we show constraints on the $\alpha_s$--$T_{\rm vir}$ (left) and $\beta_s$--$T_{\rm vir}$ (right) planes with the other running parameter fixed as
$\beta_s = 0.025$ (left) and $\alpha_s = 0.009$ (right) which corresponds to the central values derived from Planck \cite{Ade:2015lrj}.
As seen from the left panel, the constraint on $\alpha_s$ is not much affected by the value of $T_{\rm vir}$. This also suggests that 
other cosmological parameters such as $f_\ast$ and $\zeta_X$ do not affect the constraint on $\alpha_s$ much. 
On the other hand, the constraint on $\beta_s$ is much affected by the value of $T_{\rm vir}$ assumed. As discussed above, 
larger value of $T_{\rm vir}$ shifts the absorption line to lower redshift, and hence larger $T_{\rm vir}$ allows more positive values of $\beta_s$. 
However, we should note that when $T_{\rm vir} \lesssim 5 \times 10^4~{\rm K}$, the EDGES result indicates $\beta_s \lesssim 0.012$ which means that 
the Planck allowed region may be disfavored for $T_{\rm vir} \lesssim 5 \times 10^4~{\rm K}$.

Although the constraints on $\beta_s$ and $T_{\rm vir}$ are degenerate from EDGES data, 
measurements of faint galaxies at high-$z$ would be useful  to resolve the degeneracy between $\beta_s$ and  $T_{\rm vir}$. For example, based on massive data of Hubble Space Telescope (HST), faint end magnitude of UV luminosity function is $-18$ at $z=8$ \cite{Bouwens2015}, which corresponds to  $T_{\rm vir} \simeq 2\times 10^5 \rm [K]$ \cite{Greig2015}. Furthermore, the gravitational lensing method can help to find much fainter galaxies \cite{Livermore2017}. Meanwhile, the James Webb Space Telescope (JWST) is supposed to lunch in 2021 and expected to  find fainter galaxies \cite{Gardner2006}. Thus, $T_{\rm vir}$ will be constrained  by combining the gravitational lensing method and the JWST. In distant future, $T_{\rm vir}=10^4 {\rm [K]}$, corresponding to atomic cooling, might be constrained by next generation telescopes such as the High Definition Space Telescope \cite{Dalcanton2015} and the Advanced Technology Large Aperture Space Telescope\cite{Postman2010}.

Although we only show the figures in the  $\alpha_s$--$T_{\rm vir}$  and $\beta_s$--$T_{\rm vir}$ planes, 
one can also infer the effects of other astrophysical parameters on the constraints on the running parameters.
For example, due to the degeneracy between $T_{\rm vir}$ and $f_\ast$, if $f_\ast$ is assumed to be smaller than $f_\ast = 0.05$ which is the fiducial value in our analysis, 
the  allowed region of $\beta_s$ would  be shifted to the direction of larger $\beta_s$.

Regarding $\zeta_X$,  since  the number of X-ray photons emitted from stars $\zeta_X$ affects gas heating, but not the WF coupling, 
the evolution of $T_b$ at high redshift is not much affected by the change in $\zeta_X$ as seen from the bottom panel of Fig.~\ref{fig:GSastro}.
At lower redshift, the change of $\zeta_X$ gives a different efficiency of gas heating, which affects the lower part of the absorption trough.
For example, by assuming a smaller value for $\zeta_X$, $T_b$ gets cooler due to less heating, which shifts the peak redshift of the absorption line to a lower one.
Therefore, with smaller $\zeta_X$, only the lower bound of the constraint on $\alpha_s$ and $\beta_s$ would be  shifted to more larger values, while the upper bounds are unaffected.
On the other hand, powerful heating due to the higher value of $\zeta_X$ can increase $T_b$ at lower redshifts.
We also note that $\zeta_X$ affects the ionization history,  it would give a very important effect when we look at lower redshift as $z \lesssim 10$, 
which however  is not  important in the redshift range we consider in this paper.

Here we should also mention that there are other parameters which affect the X-ray heating,  such as the minimum energy of X-ray heating, $E_0$, and the spectral index of X-ray radiation, $\alpha_{X}$. 
Effects of these parameters on the behavior of 21cm global signal have been discussed in \cite{2018Greg},
from which one can see that their impact on the evolution of $T_b$  is similar to the one by $\zeta_X$, but weaker than $\zeta_X$. Thus, 
we do not show the dependence of 21cm global signal on $E_0$ and $\alpha_{X}$ here.

Before closing this section, we make a comment on a possibility of constraining primordial power spectrum 
from the 21cm global signal at lower redshifts.
The running parameters can also be constrained via the information of the signal at lower redshifts, especially from the argument of the ionization history. 
However, the evolution of ionization fraction depends on other additional parameters such as the escape fraction of ionizing photon, ionization efficiency, 
maximum mean free path of ionizing photon and $T_{\rm vir}$ during the EoR era\footnote{
For a work on constraints on ionization history at low-redshift motivated by  the EDGES result, see \cite{Witte:2018itc}.
}. 
It would be worth investigating an attainable constraint on the running parameters by using the information at such lower redshifts. 
However, we need to take account of the effect of various astrophysical parameters mentioned above, which is 
computationally very demanding. We leave this for future work. 

%%%%%%%%%%%%%%%%%%%%%%%%%%%%%%%%%%

%%%%%%%%%%%%%%%%%%%%%%%%%%%%%%%%%%%%%%%%%%%%%%%%%%
\section{Conclusion} \label{sec:conclusion}
%%%%%%%%%%%%%%%%%%%%%%%%%%%%%%%%%%%%%%%%%%%%%%%%%%

We have investigated the impact of the recent EDGES result on primordial power spectrum,
particularly focusing on the running parameters $\alpha_s$ and $\beta_s$. EDGES has detected 
the absorption line at the frequency corresponding to $z=17.2$ and also  
indicates that the absorption line should not appear in the redshift range of $z<14$ and   $z >22$.
Since large values of the runnings $\alpha_s$ and $\beta_s$ directly enhance/suppress the matter power on small scales, 
the process of the structure formation is much affected, which changes the position of the absorption trough of the 21cm global signal.
By requiring that the absorption line should appear at the frequency consistent with the EDGES result,
we have obtained the bounds on the running parameters $\alpha_s$ and $\beta_s$. In particular, for some values of astrophysical parameters such as 
$T_{\rm vir}, f_\ast$ and $\zeta_X$, the parameter space of $\alpha_s$ and $\beta_s$ allowed by Planck can be excluded in the light of the EDGES result. 

Since the absorption line detected by EDGES was somewhat unexpected, and hence 
further analysis with other instruments will be awaited to draw a more rigorous conclusion 
regarding constraints on primordial power spectrum.
Especially, more systematic analysis for  foreground removal would be inevitable. 
However, as our analysis demonstrates that the 21cm global signal is very powerful in constraining  primordial power spectrum,
 future observational/theoretical studies of the 21cm global signal would bring us more insight to understand 
the primordial Universe.

\bigskip
{
{\sl Note added:} 
After the submission of this manuscript, Planck made the final data release \cite{Aghanim:2018eyx,Akrami:2018odb},
in which constraints on the running parameters are given by 
\begin{equation}
%\label{ }
\alpha_s =  0.013 \pm 0.012\, (0.002 \pm 0.010) \, , 
\qquad
\beta_s = 0.022 \pm 0.012\, (0.010 \pm 0.013) \, ,
\end{equation}
from the data set of Planck 2018 TT(TT, TE, EE)+lowE+lensing  \cite{Akrami:2018odb}. 
Although these values are slightly different from Planck 2015 result which we mentioned in the main text, 
our arguments remain unchanged.
}

%%%%%%%%%%%%%%%%%%%%%%%%%%%%%%%%%%%
\section*{Acknowledgments}
%%%%%%%%%%%%%%%%%%%%%%%%%%%%%%%%%%%
We would like to thank Bradley Greig for helpful comments on the performance of the 21cmFAST.
This work is partially supported by JSPS KAKENHI Grant Number 15K05084~(TT),  16H05999~(KT), 16J01585~(SY),  17H01110~(KT), 17H01131~(TT), 
MEXT KAKENHI Grant Number 15H05888~(TT), 15H05896~(KT),  
and Bilateral Joint Research Projects of JSPS~(KT).

%%%%%%%%%%%%%%%%%%%%%%%%%%%%%%%%%%%%%%%%%%%%%%%%%%

%%%%%%%%%%%%%%%%%%%%%%%%%%%%%%%%%%%%%%%%%%%%%%%%%%

\end{document}